\advance\voffset by 1mm
\advance\hoffset by -7mm
\documentstyle[12pt]{article}
\def\nlni{\par\ifvmode\removelastskip\fi\vskip\baselineskip\noindent}
\newenvironment{remark}{\nlni\begingroup\it Remark. \rm}{
                        \endgroup\vskip\baselineskip}
\makeatletter
\@addtoreset{equation}{section}
\makeatother

\newcommand{\bstar}{\begin{petit} \noindent {\Large $\star$} }
\newcommand{\estar}{\end{petit}}
\def\ftoday{{\sl  \number\day \space\ifcase\month
\or Janvier\or F\'evrier\or Mars\or avril\or Mai
\or Juin\or Juillet\or Ao\^ut\or Septembre\or Octobre
\or Novembre \or D\'ecembre\fi
\space  \number\year}}
\newcommand{\journal}[4]{{\em #1~}#2\,(19#3)\,#4;}

\newcommand{\hpa}{\journal {Helv. Phys. Acta}}

\newcommand{\ijmp}{\journal {Int. J. Mod. Phys.}}

\newcommand{\cmp}{\journal {Commun. Math. Phys.}}

\newcommand{\np}{\journal {Nucl. Phys.}}
\newcommand{\pl}{\journal {Phys. Lett.}}

\newcommand{\prep}{\journal {Phys. Rep.}}

\newcommand{\nc}{\journal {Nuovo Cimento}}

           \newcommand{\G}{\Gamma}
\renewcommand{\d}{\delta}         \newcommand{\D}{\Delta}
\newcommand{\e}{\varepsilon}

\newcommand{\m}{\mu}
\newcommand{\n}{\nu}
\newcommand{\om}{\omega}         
             
\newcommand{\r}{\rho}
\newcommand{\s}{\sigma}           \renewcommand{\S}{\Sigma}

\newcommand{\vf}{{\varphi}}
\newcommand{\EE}{{\cal E}}
\newcommand{\FF}{{\cal F}}
\newcommand{\GG}{{\cal G}}

\newcommand{\LL}{{\cal L}}

\newcommand{\OO}{{\cal O}}

\newcommand{\SS}{{\cal S}}

\newcommand{\WW}{{\cal W}}

\newcommand{\Dleft}[1]{
{\raise .7ex \hbox{${\raise -.7ex \hbox{$#1$}}^{^{\kern-7pt \leftarrow}}$}}
}

\newcommand{\na}{\nabla}

\newcommand{\Sla}{\raise.15ex\hbox{$/$}\kern -.70em D}
\def\h{\hbar}

\newcommand{\lp}{\left(}\newcommand{\rp}{\right)}
\newcommand{\lc}{\left[}\newcommand{\rc}{\right]}
\newcommand{\lac}{\left\{}\newcommand{\rac}{\right\}}

\newcommand{\tr}{{\rm {Tr} \,}}
\newcommand{\half}{\frac 1 2}
\newcommand{\pa}{\partial}

\newcommand{\dfud}[2]{{\displaystyle{\frac{\delta #1}{\delta #2}}}}
\newcommand{\dfrac}[2]{{\displaystyle{\frac{#1}{#2}}}}
\newcommand{\dsum}[2]{\displaystyle{\sum_{#1}^{#2}}}
\newcommand{\dprod}[2]{\displaystyle{\prod_{#1}^{#2}}}
\newcommand{\dint}{\displaystyle{\int}}

\newcommand{\sla}{\raise.15ex\hbox{$/$}\kern -.57em}
\newcommand{\twiddle}{\lower.9ex\rlap{$\kern -.1em\scriptstyle\sim$}}

\newcommand{\vev}[1]{\left\langle {#1}\right\rangle}
\newcommand{\es}{\\[3mm]}
\newcommand{\equ}[1]{(\ref{#1})}
\newcommand{\eq}{\begin{equation}}
\newcommand{\eqn}[1]{\label{#1}\end{equation}}
\newcommand{\eea}{\end{eqnarray}}
\newcommand{\eqa}{\begin{eqnarray}}
\newcommand{\eqan}[1]{\label{#1}\end{eqnarray}}
\newcommand{\ba}{\begin{array}}
\newcommand{\ea}{\end{array}}
\newcommand{\eqac}{\begin{equation}\begin{array}{rcl}}
\newcommand{\eqacn}[1]{\end{array}\label{#1}\end{equation}}

\newcommand{\bc}{{\bar c}}
\newcommand{\bGG}{{\overline \GG}}
\newcommand{\txint}{\dint d^3x\;}
\renewcommand{\ss}{{\SS_S}}

\newcommand{\rt}{{\tilde\r}}\newcommand{\st}{{\tilde\s}}

\newcommand{\rh}{{\hat\rho}}

\newcommand{\ws}{\WW^{\rm S}}
\newcommand{\wt}{\WW^{\rm T}}
\newcommand{\wte}[1]{{\WW^{\rm T}_{(#1)}}}
\newcommand{\wsx}[1]{{\WW^{\rm S}_{(#1)}}}

\newcommand{\cb}{\bar c}

\begin{document}
%**********************************************************
%***************************************************************
%PAGE DE TITRE
%*********************************************************
%\advance\voffset by -2cm

\begin{center}
{\LARGE {\sc
            {\bf UNIVERSIT\'E DE GEN\`EVE}}}
{ \raisebox{0mm}{{\small SCHOLA GENEVENSIS MDLIX}} }
\vspace{72mm}

{\LARGE On the Role of Vector Supersymmetry \\[2mm]
  in Topological Field Theory$^1$} \vspace{2mm}

%***************************************************************
\vspace{9mm}

{\large O. Piguet$^2$}\vspace{3mm}

{\small D\'epartement de Physique Th\'eorique,
                  Universit\'e de Gen\`eve}
{\small 24, quai Ernest Ansermet,
                      CH-1211 Gen\`eve 4 (Switzerland)}
\vspace{10mm}

{{\normalsize \begin{tabular}{l}
                              {\bf  UGVA---DPT 1995/02--880}\\
                              {\bf hep-th/9502033}
                         \end{tabular}   }}
\end{center}
%************************************************************************
\footnotetext[1]{Talk given at the
``XV Encontro Nacional de F\'\i sica de
Part\'\i culas e Campos'', Angra dos Reis, Brazil, October 1994.}
\footnotetext[2]{Supported in part
                 by the Swiss National Science Foundation.}
\setcounter{footnote}{2}
%*********************************************************************
\thispagestyle{empty}
\vfill
\pagebreak
%\advance\voffset by 2cm
%***************************************************************
\null
\vspace{6cm}

\centerline{\Large{\bf Abstract}}\vspace{10mm}

Vector supersymmetry is typical of topological field theory. Its role
in the construction of gauge invariant quantities is explained,
as well as its role in the cancellation of the ultraviolet
divergences. The example of the Chern-Simons theory in three
dimensions is treated in details.
\noindent
\setcounter{page}{0}
\thispagestyle{empty}
\vfill\eject

%********************************************************
%***********************************************************************
\section{Introduction}
A main characteristics of
topological gauge theories (see~\cite{bbrt} for a general
review) is their lack of local observables. In particular there
is no energy-momentum tensor because of their metric independence
and, moreover, the gauge field configurations satisfying the
equations of motion are pure gauges. Observables are
integrated gauge invariant
objects like Wilson loops, or currents localized on the
boundary, if any, of the base manifold. Other important physical,
gauge invariant
quantities are the possible anomalies and the counterterms which
correspond to the possible
renormalization of the coupling constants. We shall deal here
with the latter category of gauge invariant objects, which are
represented by space-time integrals.

The aim of this talk is to show how a supersymmetry generated by
a space-time vector valued supercharge happens to hold, and to
show that this supersymmetry generator may be used in a
a natural way for the explicit construction of the gauge
invariant objects considered. Its role in the
complete cancellation of the ultraviolet divergences in
topological theories will also be explained.

In order to keep the argument as simple and self-contained as
possible, I shall concentrate on the 3-dimensional Chern-Simons
theory.  Sect. 2 will be devoted to the description of the
model, of its gauge fixing and to its BRS invariance. Sect. 3
will present some simple facts about the BRS cohomology which
 will be needed later on.
Vector supersymmetry will be introduced in
Sect. 4 and the solution of the problem in Sect.5. Some
conclusions are proposed at the end.

The notations are those of~\cite{livre}. Generally the
statements are given without proof.
The proofs  may be found in~\cite{livre},
unless they are well-established or an explicit reference  is
given.
%***********************************************************************
\section{The $D=3$ Chern-Simons model}\label{Chern-Simons}
\subsection{The Chern-Simons action}
The Chern-Simons term\cite{schwarz,witten}
in three space-time dimensions is given by the
gauge invariant action
\eq
 S_{\rm CS} := - \frac{k}{4\pi} \txint \e^{\m\n\r} \tr
    \lp A_\m \pa_\n A_\r - \dfrac{2i}{3} A_\m A_\n A_\r \rp \ .
\eqn{chern-sim-action}
The number $k$  plays the role of the inverse of the coupling
constant\footnote{The parameter $k$ is actually quantized\cite{witten},
but this is irrelevant in perturbation theory, which is a (formal)
expansion at zero coupling, hence at infinite $k$.}
and $A_\m$ is
a Lie algebra valued gauge field, as well as
all the  fields we shall introduce
later on. Denoting such a generic field by $\vf$, we define
\eq
\vf(x) := \vf^a(x)\tau_a\ ,
\eqn{gauge-field}
where the matrices $\tau$ are the generators of the group and obey
\eq
[\tau_a,\tau_b]=if_{abc}\tau_c\ ,\qquad \tr (\tau_a\tau_b) = \d_{ab}\ .
\eqn{tau-matrices-alg}
%This means that the field $\vf$
%belongs to the adjoint representation of the gauge group.
The gauge transformations read
\eq
\d A_\m(x) = \pa_\m\om(x) + i[\om(x),A_\m(x)] =: D_\m\om(x)\ ,
\eqn{gauge-transf}
with  $\om:=\om^a\tau_a$.
The gauge group is chosen to be simple and compact. These
transformations change the integrand of the action
\equ{chern-sim-action} into a
total derivative, leaving thus the action invariant if there are no
boundary contributions and if the topology is trivial~\cite{witten}.

Expression \equ{chern-sim-action}, although
referred here to the flat Euclidean space-time, has an intrinsic
geometrical meaning and can be defined on an arbitrary
three dimensional manifold $\EE$ in a natural way.
Indeed one remarks that the
Chern-Simons action \equ{chern-sim-action} being the integral of
a 3-form, does not depend on the metric
$g_{\m\n}$ which one may introduce on $\EE$.
Explicitly, defining the gauge connection 1-form
\eq
A:=A_\m dx^\m ,
\eqn{connection-form}
one can write the Chern-Simons action as
\eq
 S_{\rm CS} = - \frac{k}{4\pi} \dint_\EE  \tr
    \lp A\;dA - \dfrac{2i}{3} A^3 \rp \ ,
\eqn{chern-sim-action-form}
the wedge symbol $\wedge$ for the multiplication of forms being
omitted.

The main consequence of the metric independence  is the vanishing of the
energy-mom\-en\-tum tensor associated to the Chern-Simons action:
\eq
T_{\rm CS}^{\m\n} = \dfud{S_{\rm CS}}{g_{\m\n}} = 0\ .
\eqn{e-m-tensor-cs}
Another important feature of the theory is the peculiar form of the
gauge field equations:
\eq
F_{\m\n}:= \pa_\m A_\n - \pa_\n A_\m - i[A_\m,A_\n] = 0\ ,
\eqn{field-str-cs}
which means that
there is also no field strength. Thus there is no
local gauge invariant quantity at all, in other words there are
no local observables. But this
remark does not imply that the present theory is physically empty. Two
types of observables may be defined. The first one is provided by the
Wilson loops, that is by the gauge invariant quantities associated to
the closed loops $C$:
\eq
\tr\;{\rm P}\;\exp{i\oint_C A_\m dx^\m}\ ,
\eqn{wilson-loop}
where the symbol ${\rm P}$ means the ``path ordered product''.
The second type of observables are present in case the
3-manifold possesses a 2-dimensional boundary : there exists then a set
of  conserved chiral current which are localized on the boundary
and which generate an infinite dimensional
algebra of conserved
charges~\cite{moore,bc-boundary,emery-piguet}.
%********************************************************************
\subsection{Gauge Fixing}
We shall fix the gauge \`a la
Landau, adding to the
Chern-Simons action \equ{chern-sim-action} the term
\eq
S_{\rm gf} := \tr \txint \lp B\pa^\m A_\m
                -\bc \pa^\m(\pa_\m c + i[c,A_\m]) \rp \ ,
\eqn{CS-land-gauge-fix}
where the Lie algebra valued fields $B$, $c$ and $\cb$
are a Lagrange multiplier, the ghost
and the antighost, respectively.

The gauge-fixed action
\eq
    {\tilde S} := S_{\rm CS} + S_{\rm gf} \ ,
\eqn{CS-gauge-fixed}
is invariant under the nilpotent BRS transformations
\eq\ba{ll}
s A_\m = D_\m c:= \pa_\m c + i[c,A_\m]   \ ,
  \qquad sc = ic^2 \ , \es    s\bc = B\ ,   \qquad sB=0 \ .
\ea\eqn{CS-g-brs}
The gauge fixing part of the action of course depends on the metric,
chosen here as the flat Euclidean
one\footnote{See~\cite{l-p-curved} for the generalization
to a curved manifold.}
$\d_{\m\n}$.
\begin{remark}
The gauge fixing part of the action is $s$-exact, i.e. it is a
BRS variation:
\eq
S_{\rm gf} = s  \tr \txint  \cb\pa^\m A_\m\ .
\eqn{s-gaugefixing}
\end{remark}
%*************************************************************
\subsection{The Slavnov-Taylor Identity}
The BRS variations of most of the fields being nonlinear, one has
to couple them to external fields in order to control their
possible renormalizations. We thus add to the action the piece
\eq
  S_{\rm ext} := \txint \tr \lp
         \r^{\m} D_\m c + i\s c^2 \rp \ ,
\eqn{CS-external}
so that the total classical action
\eq
 S(A,c,\cb,B,\r,\s) := {\tilde S} + S_{\rm ext}
\eqn{CS-total-action}
obeys the Slavnov-Taylor identity
\eq
\SS(S) := \txint \tr \lp \dfud{S}{\r^\m}\dfud{S}{A_\m}
    + \dfud{S}{\s}\dfud{S}{c}   + B\dfud{S}{\bc} \rp = 0 \
\eqn{CS-slavnov-taylor}
which plays the role of the Ward identity associated to the BRS
invariance.
\begin{remark}
Here, at the classical level, the Slavnov-Taylor identity simply
expresses the BRS invariance of the action. At the quantum level,
the classical action is replaced by the
vertex functional\footnote{In perturbation theory considered as a
formal power expansion in $\h$, i.e.
in the number of loops of the Feynman graphs, the zeroth order -- the
classical theory -- coincides with the tree graph approximation.}
\eq
\G(A,c,\cb,B,\r\s) =
S(A,c,\cb,B,\r\s)+O(\h)\ ,
\eqn{vertex-functional}
i.e. by the generating functional of the
1-particule irreducible (1PI) Green functions:
\[
\left.\dfrac{\d^n\G}{\d\vf(x_1)\cdots\d\vf(x_N)}\right\vert_{\vf=0}
=\vev{0|T(\vf(x_1)\cdots\vf(x_N))|0}_{1PI} \ ,
\]
where $\vf$ stands for any field. Then the Slavnov-Taylor
identity \equ{CS-slavnov-taylor} has to hold with $S$ replaced by
$\G$.
\end{remark}
The Slavnov-Taylor identity thus expresses the BRS invariance in
a functional form or, equivalently, in the form of identities
between Green functions.

The gauge fixing condition may also be expressed in a functional
form:
\eq
   \dfud{S}{B} = \pa A  \ ,
\eqn{gauge-cond}
Applying the functional operator $\d/\d B$ to the Slavnov-Taylor
operator defined by \equ{CS-slavnov-taylor}
yields the identity (true for any functional $\FF$)
\eq
\dfud{}{B}\SS(\FF) = \SS_\FF \dfud{\FF}{B} + \dfud{\FF}{\cb}\ ,
\eqn{com-Slavnov-gauge-cond}
where we have defined the ``linearized'' Slavnov-Taylor operator
\eq
\SS_\FF := \tr\txint\lp \tr\dfud{\FF}{\r^\m}\dfud{}{A_\m}
            + \dfud{\FF}{A_\m}\dfud{}{\r^\m}
            + \dfud{\FF}{\s}\dfud{}{c}
            + \dfud{\FF}{c}\dfud{}{\s}
            +  B\dfud{}{\bar c}\rp\ .
\eqn{slavnov-lin-op}
Applying the identity \equ{com-Slavnov-gauge-cond}
for $\FF=S$ and making use of the Slavnov-Taylor
identity\equ{CS-slavnov-taylor}
and of the gauge condition \equ{gauge-cond} yields the ``ghost
equation''
\eq
\GG\G = 0\ ,\qquad{\rm with}\quad
       \GG:=\dfud{}{\bc}+ \pa^\m\dfud{}{\r^\m} \ .
\eqn{ghost-eq}
The latter implies that the antighost $\cb$ and the external field $\r$
appear only in the combination
\eq
\rh^\mu = \r^\m + \pa^\m\bc \ .
\eqn{hat-rho}
\vfill\eject
%****************************************************************
\subsection{Stability}\label{stabilite}
The linearized Slavnov-Taylor
operator \equ{slavnov-lin-op} may be interpreted as the ``derivative at
the point $\FF$'' of the nonlinear map $\SS$,
defined by \equ{CS-slavnov-taylor}, from the space of the field
functionals into the complex numbers.
For $\FF$ equal to the action $S$, it can be shown
to be nilpotent as a consequence the Slavnov-Taylor
identity:
\eq
\ss^2=0\ ,
\eqn{nilpot-ss}
and its action on the fields reads
\eq\ba{l}
\ss\vf = s\vf\ ,\quad \vf=A,c,\cb,B\ ,\es
\ss\r^\m = \dfud{S}{A_\m}\ ,\es
\ss\s = \dfud{S}{c} \ .
\ea\eqn{lin-slavnov-action}
One sees that it coincides with the BRS operator $s$ when applied to the
dynamical fields, and yields equations of motion if applied to the
external fields.

The operator $\ss$ allows one to characterize the stability of the theory
defined by the Slavnov-Taylor identity and by the gauge condition.
This means the following. Given an action $S$ obeying the equations
\equ{CS-slavnov-taylor} and \equ{gauge-cond}, let us ask for the most
general solution $S'$ in the neighbourhood of $S$.
Writing
\eq
S'=S+\e\D\ ,
\eqn{perturbation}
with $\e$ ``small'', we find that the
perturbation
$\D$ has to obey the two conditions
\eq
\ss\D=0
\eqn{pert-brs-inv}
and
\eq
\dfud{\D}{B}=0\ .
\eqn{pert-gauge-cond}
Due to the ghost equation \equ{ghost-eq}, it must of course obey the
further  condition
\eq
\GG\D=0\ ,
\eqn{pert-ghost}
which however is not independent due to the commutation rule
\eq
\lc \dfud{}{B},\ss \rc = \GG\ .
\eqn{com-gaugecond-ss}
A model given by an action $S$ is ``stable'' if the most general
perturbation $\D$ can be obtained by a redefinition of the parameters
and of the field variables.

It is very important to distinguish the ``physical'' perturbations from
the ``unphysical'' ones. The latter perturbations are those which
correspond to a mere redefinition of the fields: they indeed don't
affect the physical outcome of a theory, e.g. the scattering  amplitudes
in a usual gauge theory. It turns out that the unphysical
$\ss$-invariant perturbations are those that take the form of a
$\ss$-variation:
\eq
\D_{\rm unphys} = \ss\hat\D\ .
\eqn{unphys-pert}
Perturbations corresponding to the redefinition of gauge parameters are
unphysical, too, and have the same form.

On the other hand, the physical perturbations correspond to the
redefinition of physical parameters like the coupling constants and the
masses. They are characterized by the property:
\eq
\ss\D_{\rm phys} = 0\ ,\quad\mbox{but }\
 \D_{\rm phys}\not= \ss\hat\D
 \ \mbox{ for any }\ \hat\D\ .
\eqn{phys-pert}
Solving \equ{pert-brs-inv} with a nilpotent operator $\ss$
is thus a problem of cohomology. The next section will deal with
this.
%********************************************************************
\section{Cohomology and Descent Equations}
\subsection{Cohomology and Observables}
In ordinary gauge theories the observables are defined, at the classical
level, as gauge invariant local polynomial of the
{\it physical fields} (gauge
and matter fields). ``Local'' from now on will
mean that these polynomials are made of
products of the fields and of their derivatives at a common
space-time point $x$. They may integrated (e.g. Wilson loops) or not
(e.g. currents).
In the quantum theory, defined e.g. by the renormalized
perturbation theory, local field polynomials $P$ become local ``operators'',
or, in the framework of the Green functions, local ``insertions''
which may be represented by the generating functional
\eq
P\cdot\G = P+O(\h)
\eqn{insertions}
of the 1-particule irreducible Green functions
\[ \vev{0|T\lp P\;\vf(x_1)\vf(x_2)\cdots\rp|0}_{1PI} \ . \]
The right-hand-side of \equ{insertions}
expresses the fact that the tree graph approximation
corresponds to the classical approximation.
But now $P$ may depend on  unphysical fields
such as the ghost fields and the external fields.

The quantum version of an observable is then defined as an insertion
given by a
quantum extension of a local polynomial $\OO$ of the {\it physical and
unphysical fields}, which is BRS-invariant
but not a BRS variation (the linearized Slavnov-Taylor
operator \equ{slavnov-lin-op} for $\FF=\S$ must be used):
\eq
\ss\OO=0\ ,\quad\mbox{but }\
\OO\not=\ss\hat\OO \ \mbox{ for any }\ \hat\OO\ .
\eqn{ss-cohomology}
In other terms, quantum observables are defined as cohomology classes of
the ``coboundary operator'' $\ss$ in the space of the local
functionals.

We have seen that topological gauge theories such as the
Chern-Simons theory
do not possess $x$-dependent local observables. But they possess
quantities which are gauge invariant in the sense defined above, i.e.
defined by the cohomology of $\SS$. We have already
encountered the example of the nontrivial BRS-invariant
perturbations of
the action in Subsect.~\ref{stabilite}. Another important example
is that of the gauge  anomalies\footnote{Their
absence in the Chern-Simons theory
was actually shown in ref.~\cite{blasi-collina}.}.
%*********************************************************
\subsection{The Descent Equations}
The structure we shall discuss here applies to the gauge
invariant quantities which are expressed as space-time integrals.
But, in order to keep simplicity, we shall
consider in more details only the case of the perturbation
$\D$ introduced in Subsect.~\ref{stabilite}.
With
\eq
\D= \txint Q^0(x)\ ,
\eqn{q0}
It follows from the BRS invariance condition \equ{pert-brs-inv}
that the BRS variation of the
integrand must be a total derivative
\footnote{The exponents give the ghost number. By
convention the ghost number of $c$ is equal to 1.}:
\eq
\ss Q^0 = \pa_\m Q^{1\m}\ .
\eqn{q1}
The conditions \equ{pert-gauge-cond} and
\equ{pert-ghost} imply that $\D$ -- as well as all
the quantities which we shall derive from it -- do not depend
on the Lagrange multiplier $B$ and depend on the antighost $\cb$
and the external field $\r^\m$ only through the combination
\equ{hat-rho}. The relevant variables are therefore the fields
$A_\m$, $c$, $\rh^\m$ and $\s$.

Applying $\ss$ to \equ{q1} and using
the nilpotency of $\ss$ now imply that the variation of
 $Q^{1\m}$ is the total derivative of an antisymmetric tensor:
\eq
\ss Q^{1\m} = \pa_\n Q^{2[\m\n]}\ .
\eqn{q2}
Repeating the argument twice leads to
\eq
\ss Q^{2[\m\n]} = \pa_\r Q^{3[\m\n\r]} \ ,\quad
 \ss Q^{3[\m\n\r]} = 0  \ .
\eqn{q3}
The process stops here because the rank of an antisymmetric
tensor  is bounded by the space-time dimension.

In the notation of differential forms, these ``descent
equations'' read
\eq\ba{l}
\ss\om^0_3 = d\om^1_2\ ,\es
\ss\om^1_2 = d\om^2_1\ ,\es
\ss\om^2_1 = d\om^3_0\ ,\es
\ss\om^3_0 = 0 \ ,
\ea\eqn{desc-eq}
with
\eq
\D= \dint \om^0_3\ .
\eqn{om03}
Here, $d$ is  the exterior derivative (with $d^2=0$), and the
forms $\om^g_{3-g}$ of ghost number $g$ and degree $3-g$ are
defined by:
\eq\ba{l}
\om^0_3 = \dfrac{1}{3!} \e_{\m\n\r}Q^0 dx^\m dx^\n dx^\r\ ,\es
\om^1_2 = \dfrac{1}{2} \e_{\m\n\r}Q^{1\r} dx^\m dx^\n\ ,\es
\om^2_1 = \dfrac{1}{2} \e_{\m\n\r}Q^{2\n\r} dx^\m\ ,\es
\om^3_0 = \dfrac{1}{3!} \e_{\m\n\r}Q^{3\m\n\r}  \ .
\ea\eqn{def-omega}
\begin{remark}
The formalism of the differential forms is the natural one in the
present context of a topological theory. Indeed, \equ{om03} as
well as the descent equations \equ{desc-eq} never involve the
metric, but only the differentiable structure of the manifold. To
the contrary the metric enters explicitly, although in a
spurious way, in
the notation in terms of the tensors $Q$, the latter
being the Hodge duals of the forms $\om$.
\end{remark}
%**********************************************************
%--------------------------------------------------
\subsection{Solution: Existence and
Uniqueness}\label{exist and unique}
In order to solve the descent equations, let us begin by the last
and simplest of them. It is clear that, since $\om^3_0$ is a 0-form
of ghost number 3, its most general invariant expression is
given by
\eq
\om^3_0 = x\dfrac{i}{3} \tr c^3\ ,
\eqn{om30}
with $x$ an arbitrary coefficient. In order to find the solution
for the other forms, and in particular for $\om^0_3$  yielding
the quantity $\D$ \equ{om03}
of interest, one has to climb up the descent
equations. Each step represents a cohomology problem, but for
nonintegrated objects. This cohomology can be shown to consists
only of invariant
0-forms made with the ghost field $c$ alone, without
derivative\footnote{This is very peculiar to the topological
theories. In ordinary Yang-Mills theories the cohomology depends
on the invariantsdraft constructed with the ghost $c$ and on
those made with the Yang-Mills strength $F$ as
well~\cite{dixon,dubois,brandt}.}.
This result was just used above in order to solve
the last descent equation. However, since the next steps involve
forms of higher degree, the cohomology is then trivial. As it can be
shown, this
implies that, once the 0-form $\om^3_0$, i.e. the coefficient
$x$, is given, the general solution for the 3-form $\om^0_3$ is
{\it unique} up to the BRS variation of a an arbitrary 3-form
 and the exterior derivative of an arbitrary 2-form:
\eq
\om^0_3 = {\bar\om}^0_3 + \ss{\hat\om}^{-1}_3
                                  + d{\hat\om}^0_2\ ,
\eqn{gen-sol}
where ${\bar\om}^0_3$ is a particular solution. We will show in
the next section how a supersymmetry generated by a vector
operator indeed gives a particular solution, and what this
implies.
%********************************************************************
\section{The Vector Supersymmetry}
Since the action depends on the metric only through the gauge fixing
part \equ{CS-land-gauge-fix}, which is a BRS variation, the
energy-momentum tensor will be a BRS variation as well:
\eq
T^{\m\n} = \dfud{S_{\rm gf}}{g_{\m\n}} = \ss Q^{\m\n} \ .
\eqn{T=sQ}
Although the conservation of $T^{\m\n}$ does not strictly imply
that of $Q^{\m\n}$, it turns out~\cite{dlps}
that $Q^{\m\n}$ is indeed conserved\footnote{This actually happens
in every topological model studied up to now.}.
It follows that there exists a conserved
vector charge $Q_\m$ of ghost number
$-1$. This charge generates  infinitesimal
``vector supersymmetry'' transformations, displayed here
for all the fields in
regard with their BRS transformations (see \equ{lin-slavnov-action}):
\eq\ba{ll}
\ss c = ic^2 &Q_\m c = -A_\m \es
\ss A_\n = D_\n c
    \qquad&Q_\m A_\n = \dfrac{2\pi}{k} \e_{\m\n\r} \rh^\r\es
\ss \rh^\n = -\dfrac{k}{4\pi}\e^{\n\r\s}F_{\r\s} + i\{c,\h^\n\}
  \qquad&Q_\m\rh^\n = -\d_\m^\n \s  \es
\ss \s = D_\n\rh^\n + i[c,\s] \qquad&Q_\m\s =0 \es
\ss \cb = B  &Q_\m\cb=0 \es
\ss B=0      &Q_\m B = -\pa_\m\cb  \ .
\ea\eqn{transform-comp}
The variation of the action
under these supersymmetry transformations reads
\eq\ba{l}
\wsx{\xi}S := \dint\dsum{\mbox{all fields $\vf$}}{}
  \d^{\rm S}_{\xi}\vf\dfrac{S}{\d\vf} \es\phantom{\wsx{\xi}S :}
  = \D_\m^{\rm class} \; := \;
  \dint\tr\lp \dfrac{2\pi}{k}\e_{\m\n\r}\r^\n\pa^\r B
   + \r^{\tau}\pa_{\m} A_{\tau} - \s \pa_{\m} c \rp \ ,
\ea\eqn{susy-wi}
where we have denoted the infinitesimal supersymmetry
transformation of parameter $\xi^\m$ by
\eq
\d^{\rm S}_{\xi} := \xi^\m Q_\m\ .
\eqn{def-W-s}
We see that the action is actually invariant for vanishing
external fields $\r$ and $\s$. For the quantum theory, where the
action $S$ is replaced by the functional $\G$, \equ{susy-wi} is
interpretated as a (broken)Ward identity for vector
supersymmetry. The breaking $ \D_\m^{\rm class}$
vanishes when the external fields are set to zero.
Moreover, being {\it linear} in the quantized fields, it does not
need to be renormalized -- whence the superscript ``class''.
It is therefore
harmless\footnote{This would not be the case for an arbitrary
gauge fixing. One can check that, at least
among the family of the linear covariant gauges, vector
supersymmetry selects the Landau
gauge. In particular
a Feynman-type gauge would violate supersymmetry.
Some noncovariant gauges, like the axial gauge, are also
possible~\cite{blpss,blepss,emery}.}.

The geometrical aspect of the
supersymmetry transformation laws is much more apparent
if one write them -- for all the fields except $\cb$ and $B$ --
in the language of differential forms.
Introducing the forms
\eq
\rt = \half \e_{\m\n\r}\rh^\m dx^\n dx^\r\ ,\quad
\st = \dfrac{1}{3!} \s dx^\m dx^\n dx^\r ,
\eqn{forms-rho-sigma}
(the form $A$ being already defined by \equ{connection-form}),
and denoting by $i_\xi$ the interior derivative, or contraction,
with respect to the the vector $\xi^\m$,
one can write, in an obviously metric independent way,
\eq\ba{ll}
\ss c = ic^2 &\d^{\rm S}_{\xi} c = - i_\xi A\es
\ss A = dc+i\{c,A\}
    \qquad&\d^{\rm S}_{\xi} A = -\dfrac{2\pi}{k} i_\xi\rt \es
\ss \rt = \dfrac{k}{4\pi}(dA+iA^2) + i\{c,\rt\}
  \qquad&\d^{\rm S}_{\xi}\rt = -i_\xi\st  \es
\ss \st = d\rt+i\{A,\rt\} + i\{c,\st\}
\qquad&\d^{\rm S}_{\xi}\st =0 \ .
\ea\eqn{transform-forms}
Now, with the help of the properties
\eq
d^2=0\ ,\qquad \{d,i_\xi\} = \LL_\xi\ , \qquad
   \{i_\xi,i_{\xi'}\} = 0\ ,
\eqn{lie-der}
where $\LL_\xi$ denotes the Lie derivative
along the vector
$\xi$,  it is easy to check that the algebra
\eq
\ss=0\ ,\quad \lac \ss,\wsx{\xi}\rac = \wte{\xi}\ ,\quad
  \lac \wsx{\xi},\wsx{\xi'} \rac = 0\ ,
\eqn{s-algebra}
holds, where
\eq
\wte{\xi}:= \int\dsum{\mbox{all fields $\vf$}}{}
  \LL_{\xi}\vf\dfud{}{\vf}
\eqn{translation-gen}
is the generator of the space-time
translations\footnote{In the flat
space and with the constant vector fields considered here, this
Lie derivative takes the simple form of an infinitesimal translation
$\xi^\m\pa_\m$. In
a general curved manifold $\LL_\xi$ generates
the diffeomorphisms along the vector field
$\xi$, and supersymmetry transformations become
superdiffeomorphisms~\cite{l-p-curved}.}.
This shows that the BRS, supersymmetry and translation
generators $\ss$, $\wsx{\xi}$ and $\wte{\xi}$
obey a superalgebra of the Wess-Zumino type.

Let us end this section by noting that
the supersymmetry Ward identity \equ{susy-wi} imposed to the theory
yields the constraint
\eq
\wsx{\xi}\D = 0
\eqn{susy-constraint}
on the perturbation \equ{perturbation}, which has to be obeyed together
with the previous constraints \equ{pert-brs-inv},
\equ{pert-gauge-cond} and \equ{pert-ghost}.

%*********************************************************
\section{Solving the Descent Equations}
We have mentioned in Subsect.~\ref{exist and unique} that the
descent equations \equ{desc-eq}  admit a
unique solution -- modulo a total derivative and a BRS variation
-- once the bottom 0-form is given. We shall show in the
present section, on the example of the equations \equ{desc-eq},
how vector supersymmetry allows to find that solution.

This is done with the help of the
operator\footnote{A more
intrinsic definition of $\na$,
suitable for the generalizations, is provided by its action on
the individual fields (in the differential form notation):
\eq\ba{l}
\na c = -A \  , \quad \na A = -\frac{2\pi}{k}\rt \  , \quad
\na\rt = -\st \  , \quad \na\st = 0\ .
\ea\eqn{def-na-forms}}
\eq
\na := dx^\m\WW_\m\ ,
\eqn{def-na}
where we have introduced the components $\ws_\m$ of the
supersymmetry generator  \equ{susy-wi} -- and also
$\wt_\m$ for the translations -- defined by
\eq
\wsx{\xi} =: \xi^\m\ws_\m\ ,\quad \wte{\xi} =: \xi^\m\wt_\m\ .
\eqn{def-w-mu}
It is clear from the algebra \equ{s-algebra} which now reads
\eq
\ss^2=0\ ,\quad \lac \ss,\ws_\m \rac = \wt_\m\ ,\quad
  \lac \ws_\m,\ws_\n \rac = 0\ ,
\eqn{s-alg-comp}
that
\eq
\dprod{k=1}{n} \ws_{\m_k} = 0 \quad\mbox{ if } n\ge3  \ .
\eqn{susy**n}
Hence
\eq
\na^n =  0 \quad\mbox{ if } n\ge3  \   .
\eqn{nabla**n}
The algebra  \equ{s-algebra}
also implies  the commutation relation\footnote{This is
Sorella's ``decomposition of the exterior
derivative''~\cite{sorella}.}
\eq
[\na,\ss] = d\ .
\eqn{dec-ext-der}
The particular solution of the descent equations
we are looking for reads
\eq
\om^{3-p}_p = \dfrac{1}{p!}\na^p\om^3_0 \ ,\quad p = 0,\cdots,3\ .
\eqn{nabla-p}
This is readily checked using the commutation relation \equ{dec-ext-der}.
The corresponding perturbation $\D$ \equ{om03} of the action then reads
\eq
\D = \dfrac{1}{3!}\dint \na^3 \om^3_0 =
  ix\dint\tr\lp \dfrac{k}{6\pi}A^3 + \st c^2 + \rt\{A,c\} \rp\ .
\eqn{c-term}
This is only one of the BRS invariant solutions
corresponding to the same zero 0-form $\om^3_0$,
as we already know. Another possible solution
is e.g. the one which leads to
the Chern-Simons action \equ{chern-sim-action}:
\eq
\D=x \S_{\rm CS}\ .
\eqn{delta=cs}
(One can actually check that \equ{c-term} and \equ{delta=cs} differ by
a BRS variation.)
The solution \equ{delta=cs} would correspond
to a renormalization of the parameter $k$. However the
Chern-Simons action $\S_{\rm CS}$ is not supersymmetric and one has
thus to discard this solution.

On the other hand, the solution \equ{c-term} being obtained
by the triple application of the supersymmetry operator
is supersymmetric  due to
the identity \equ{susy**n} taken for $n=4$.
Moreover, it is the only supersymmetric one, as we shall check below.
But before, let us mention that, like \equ{delta=cs}, it
also corresponds to a renormalization of $k$, however combined with a
renormalization of the field amplitudes.

Let us write the general solution, which  differs from \equ{c-term}
by an arbitrary BRS variation  (restricted by having the same
quantum numbers and dimension as the action):
\[\ba{l}
\D_{\rm general} = \D + \ss\dint\tr\lp y \st c + z \rt A \rp \es
\phantom{\D} = \D
 +  \dint\tr\lp y (i\st c^2+\rt dc + i\rt\{A,c\})
   + z \lp \dfrac{k}{4\pi}(dAA+iA^3) -\rt dc  \rp \rp\ .
\ea\]
But the variational part in
the right-hand side is not supersymmetric for any ratio of the
parameters $y$ and $z$: this shows that indeed \equ{c-term} is the
unique perturbation of the action which is both BRS invariant and
supersymmetric.
%******************************************************************
\section{Ultraviolet Finiteness}
\subsection{Classical Perturbation}
Gauge theories quantized in the Landau gauge are characterized by a
remarkable property, which may be stated in words  as the
nonrenormalization of the insertions depending on the ghost field $c$
not differentiated. This property is expressed by the ``antighost
equation''~\cite{bps} which, in the present case reads
\eq
\bGG S :=  \txint \lp -i \dfud{\ }{c} + \lc \bc, \dfud{\ }{B} \rc \rp S
   =  \Delta^G_{\rm class} \; := \;
\txint \lp [\rho^\m, A_\m]  - [\sigma,c]  \rp \ ,
\eqn{antighost-eq}
where the breaking $\Delta^G_{\rm class}$, like the one of the
supersymmetry Ward identity, is linear in the quantum fields, hence
not subject to renormalization.
For the perturbation $\D$ defined in Subsect.~\ref{stabilite}, this
leads to  the further constraint
\eq
\txint  \dfud{}{c} \D = 0\ .
\eqn{ant-gh-pert}
Now, it is visible that \equ{c-term} violates this latter
constraint, and thus must be discarded. As a final result, taking into account
all the constraints dictated by the initial theory on the perturbation
$\D$ leads to the solution
\eq
\D = 0 \ .
\eqn{sol-zero}
%****************************************************************
\subsection{Counterterms Induced by the Radiative Corrections}
Up to now, we have spoken only of the classical theory, and more
specifically  of its stability. The outcome is that the
classical theory is not only stable: its parameters themselves are
completely fixed.

Going to the quantum theory, in the perturbative framework, requires
the investigation of two points.
%***********************
\subsubsection*{1) Absence of anomalies}
One has to check the validity to all orders of the Slavnov-Taylor
identity \equ{CS-slavnov-taylor}, of the
supersymmetry Ward identities \equ{susy-wi}, and as well of the
constraints given by the gauge condition, the ghost equation and the
antighost equation. In other words one has to check that all these
identities do not suffer from anomalies. It has actually been
verified~\cite{blasi-collina,dlps,livre} that algebraic consistency
forbids the appearance of such anomalies.
%****************
\subsubsection*{2) Absence of counterterms}
One has to check the stability of the theory upon the radiative
corrections: these corrections must be interpretable at
each order as a renormalization of the parameters of the theory. For
this one has to study all possible counterterms to the action. The
counterterms are of course constrained by the symmetries and identities
imposed to the Green functions. It is a very general feature of
renormalization theory~\cite{livre} that these constraints on
the counterterms are identical to
the constraints on the perturbation of the classical action we have
already spoken about. We can therefore retain our previous result
\equ{sol-zero}. This simply  means the absence of any free
counterterm -- hence
of any free parameter once those of the classical theory have been fixed.
The usual interpretation in the framework of a
renormalization via
a cut-off procedure is the ultraviolet finiteness of the theory.
One can however offer a
more physical interpretation~\cite{dlps}, namely that
the quantum theory keeps intact the {\it scale invariance} of the original
classical theory.
%**************************************************************
\section{Conclusion}
The role of vector supersymmetry, as we have seen, is twofold.

Its first role is to realize explicitly
Sorella's decomposition of the external derivative \equ{dec-ext-der},
thus providing the operator $\na$ for the construction of solutions to
the descent equations.

The second aspect of this supersymmetry is its role in the cancellation
mechanism of the ultraviolet divergences.
In fact, as we have seen in the
Chern-Simons example, supersymmetry alone is not sufficient. The
special
coupling of the ghost $c$ as expressed by the ``antighost equation''
\equ{antighost-eq} is also needed. One actually observes a
``conspiration'' between both conditions: supersymmetry selects the unique
element \equ{c-term} among the whole family of BRS invariant
counterterms, and it is just this element which is killed by the
the condition \equ{ant-gh-pert} following from the antighost equation.

Many of the results shown here for the three-dimensional Chern-Simons
theory have been generalized to a whole class of topological theories,
including the $BF$ models~\cite{abud,wallet,magg-sor,lps},
the bosonic
string ~\cite{BBS,Becchi}, four-dimensional topological Yang-Mills
theory~\cite{werneck,top-ym}, etc.
%*********************************************************************
\bigskip

\noindent{\bf Acknowledgements}:
The author would like to manifest his gratitude
to the organizers of
this conference for their kind invitation, and the
CNPq for
its financial support which made
possible the present contribution.
He also expresses his gratitude
to all his colleagues from Brazil
whom he had the opportunity to meet
for many discussions on physics and
for their warm hospitality as well. Finally, he thanks Nicola
Maggiore for his critical reading of the manuscript.
%******************d*************************************

%********************************************************
%********************************************************
\end{document}